# Consciousness is entailed by compositional learning of new causal structures in deep predictive processing systems.

V.A. Aksyuk,

NATIONAL INSTITUTE OF STANDARDS AND TECHNOLOGY,

100 BUREAU DR., GAITHERSBURG MD 20899 USA

VLADIMIR.AKSYUK@NIST.GOV

## Abstract

Machine learning algorithms have achieved superhuman performance in specific complex domains. However, learning online from few examples and compositional learning for efficient generalization across domains remain elusive. In humans, such learning includes specific declarative memory formation and is closely associated with consciousness. Predictive processing has been advanced as a principled Bayesian framework for understanding the cortex as implementing deep generative models for both sensory perception and action control. However, predictive processing offers little direct insight into fast compositional learning or of the separation between conscious and unconscious contents. Here, propose that access consciousness arises as a consequence of a particular learning mechanism operating within a predictive processing system. We extend predictive processing by adding online, single-example new structure learning via hierarchical binding of unpredicted inferences. This system learns new causes by quickly connecting together novel combinations of perceptions, which manifests as working memories that can become short- and long-term declarative memories retrievable by associative recall. The contents of such bound representations are unified yet differentiated, can be maintained by selective attention and are globally available. The proposed learning process explains contrast and masking manipulations, postdictive perceptual integration, and other paradigm cases of consciousness research. 'Phenomenal conscious experience' is how the learning system transparently models its own functioning, giving rise to perceptual illusions underlying the meta-problem of consciousness. Our proposal naturally unifies the feature binding, recurrent processing, predictive processing, and global workspace theories of consciousness.

## INTRODUCTION

Machine learning has advanced explosively in the last three decades and has enabled numerous practical applications in the areas of image and language processing and autonomous control. Deep learning and reinforcement learning(Sutton and Barto 1998) principles enabled machines that learned chess, go and computer games from scratch, achieving superhuman performance(Silver et al. 2018; Mnih et al. 2015). While these advances may have been inspired by natural systems, the overarching information processing principles in the biological brain remain poorly understood. Efficient and general in situ learning and behavioral adaptation have evolved that confer survival and reproductive advantages, justifying significant energetic and other fitness costs. In contrast, artificial learning systems are still unable to quickly learn in situ from limited data or to compositionally generalize learned perceptions and behaviors across different contexts(Kaelbling 2020)—the tasks mammals excel at. In humans, this fast compositional learning is closely linked to the formation of declarative memories, which are seemingly inseparable from conscious information processing(Kandel, Schwartz, and Jessell 2000). Thus, understanding consciousness is not only one of the preeminent intellectual challenges of our time but also promises radical practical advances for the next generation of artificial learning agents.

Since the turn of the century, the problem of human consciousness has been the focus of intense and growing experimental, theoretical and philosophical attention. While initial empirical investigations were organized around searching for the neural correlates of consciousness(Crick and Koch 1990), more recently, a variety of conceptual frameworks and theories of consciousness (ToCs)(Seth and Bayne 2022) have been advanced to provide the intellectual scaffold for orienting the empirical studies. While all leading ToCs capture important empirically supported insights, they still remain mutually incongruent, largely focusing on separate aspects of consciousness and supporting neurobiological and empirical data(Doerig, Schurger, and Herzog 2021). There is currently no conceptual approach that can unify these descriptions as aspects of a common model addressing all consciousness-relevant observations, both empirical and first-person. Ultimately, consciousness must be understood and defined as an inherent aspect of a broader unified model of cognition, which seamlessly, includes attention and affect with learning, perception and action, and describes how the perception of having a 'subjective experience' arises within the system, and how the system's overall function confers an evolutionary advantage.

The relationship between consciousness and learning plays a prominent role in several ToCs(Lamme 2006; Cleeremans et al. 2020; Cleeremans 2011; Birch, Ginsburg, and Jablonka 2020; Singer 2001), including a recent proposal relating consciousness to memory formation(Budson, Richman, and Kensinger 2022). Consciousness has been connected to the dynamic binding of known perceptual features for representing novel compound objects(von der Malsburg 1999; Singer 2001; Crick and Koch 2003; Treisman 2003), as well as to learning and dynamically forming connections among perceptual features via recurrent processing(Lamme 2006). The predictive processing framework(Hohwy and Seth 2020; Hohwy 2020; Andy Clark 2013) and active inference(K. Friston et al. 2017; K. J. Friston et al. 2010; Parr and Friston 2019) describe learning generative hierarchical relationships between causes, which encode perception and action. Finally, reinforcement learning(Montague, Dayan, and Sejnowski 1996; Schultz, Dayan, and Montague 1997; Cohen et al. 2012) provides a powerful action learning framework with an explicit connection to valence and affect. It can also be viewed as a particular case of active inference, giving a principled approach to the exploration-exploitation tradeoff. Neural implementations of these functions have been broadly discussed, with specific neural circuits proposed for predictive processing and active inference and experimentally studied for reinforcement learning.

Many general and flexible gene proliferation strategies, including not only maintaining organism homeostasis but also maximizing reproduction, are evidently being learned online and implemented by the brain. In particular, efficient declarative learning, proceeding through working memory(Kandel, Schwartz, and Jessell 2000), is directly related to consciousness – humans form declarative memories of conscious contents. This ability to dynamically compose new associations of multiple perceptions and actions immediately enables flexible and adaptive behavioral responses to novel stimuli while forming building blocks for further hierarchical compositional learning. It is therefore compelling to consider consciousness as a manifestation of a compositional learning process.

Here, we develop a conceptual proposal for the functional organization of a biological or an artificial conscious agent as a compositional learning system that combines learning by binding with predictive processing (PP) for perception and active inference and, particularly, reinforcement learning (RL) for action. The proposed learning architecture constantly posits and tests new perceptual hypotheses, attempting to find common hidden causes to reduce its largest, most persistent, and time-correlated prediction errors. Such new causes selectively bind, as their features, a small subset of all the ongoing perceptual inferences, and we argue that these sparse compositions constitute the system's conscious contents. The inherent sparseness and time selectivity of this learning-by-binding process naturally explains the apparent consciousness information processing bottlenecks in scope and time. We show how the proposed functional model entails unified yet differentiated conscious contents, short- and long-term memory formation and associative recall, attention and working memory. We explain key paradigm cases of consciousness research, including stimulus contrast, timing and masking manipulations, and sequence integration and postdiction effects. By directly incorporating predictive processing and feature binding, our proposal not only closely relates to learning in recurrent processing but also entails some of the key insights underlying the global workspace(Baars 1995; 2005; Dehaene and Changeux 2011; Mashour et al. 2020) and high-order theories(Cleeremans et al. 2020; Cleeremans 2011; Brown, Lau, and LeDoux 2019), providing common ground among the major ToCs. We address the meta-problem of consciousness(Chalmers 2018) by describing how the perceptual representations the system infers to differentiate its internal states give rise to first-person conscious perceptions of the ineffable 'what it is like' to have conscious experience. We argue that learning by binding applied to deep generative models enables sample-efficient, generalizable and compositional online learning, which is the hallmark of conscious information processing in humans and is currently lacking in artificial agents(Kaelbling 2020).

# PERCEPTION

In this section, we describe the addition of learning by binding, enabling efficient and compositional learning within the predictive processing (PP) framework. We argue that this simple additional learning mechanism accounts for the major first- and third-person observations related to conscious experience.

## Background – predictive processing framework.

Within the predictive processing (PP) framework (Figure 1a), sensory perception is conceptualized as the approximate dynamic Bayesian inference of the states of discrete causes interconnected into a deep hierarchical generative model(K. Friston and Kiebel 2009). Inference is achieved by local free energy minimization, minimizing the prediction errors for all causes throughout the model while being constrained by the sensory data at the lowest level of the hierarchy. Progressively more abstract and time-persistent causes higher in the hierarchy are inferred from, and predict, their contributing "feature" causes at the levels below. The relationships between causes and features are encoded by the prediction weight parameters connecting them. The model expresses the principle that sensory data are described by deep hierarchical bidirectional causal relationships between wholes and their features.

PP models can combine continuous-valued and categorical causes(K. J. Friston, Parr, and de Vries 2017). The simplest categorical cause has a single scalar value denoting the likelihood of it being present. For example, how likely is there being a 'cat' in the present world state? Notably, continuous-valued scalar and vector quantities, such as position, rate of change in time, and others, can be represented by sets of categorical causes approximating continuous likelihood distributions on discretized one-, two- or higher-dimensional maps, consistent with many known cortical map areas. Such maps may implement normalization, smoothing, spatial filtering and other regularization relationships.

The commonly described PP learning process gradually updates prediction weights within a fixed model topology to minimize residual local prediction errors, thereby incrementally improving the model's internal consistency and sensory data agreement. A PP model may be further constrained by domain-specific a priori learning biases to increase predictive power by reducing model complexity. These include built-in mutual constraints within subsets of causes, such as winner-take-all, or imposed time persistence for deep causes to retain their sensory data predictive power in the presence of large processing latencies. Importantly, the perceptual PP must be capable of learning dynamics at multiple timescales(K. J. Friston et al. 2017) while accounting for unequal sensory latencies, feature asynchrony and explicit time dependence. Model's causes must be able to predict short time sequences of features separated by delays and be inferred from them.

While PP already implements an information bottleneck(Still 2014) by extracting the temporally deep(K. J. Friston et al. 2017), stable and predictive sufficient statistics from the heterogeneous and rapidly varying sensory data, the PP framework for consciousness remains silent regarding the narrower bottleneck of consciousness: the apparent selection and structuring of a small subset of all ongoing perceptual inferences into a series of unified conscious contents describing perceptual episodes. The broadly parallel nature of conventional PP inference and learning does not seem to have the necessary mechanism for constructing such unified compositional (differentiated) descriptions for each present moment. Below we describe a compositional learning mechanism that not only adds efficient new structure learning to PP, but, we argue, forms the perceptual descriptions that constitute the conscious contents.

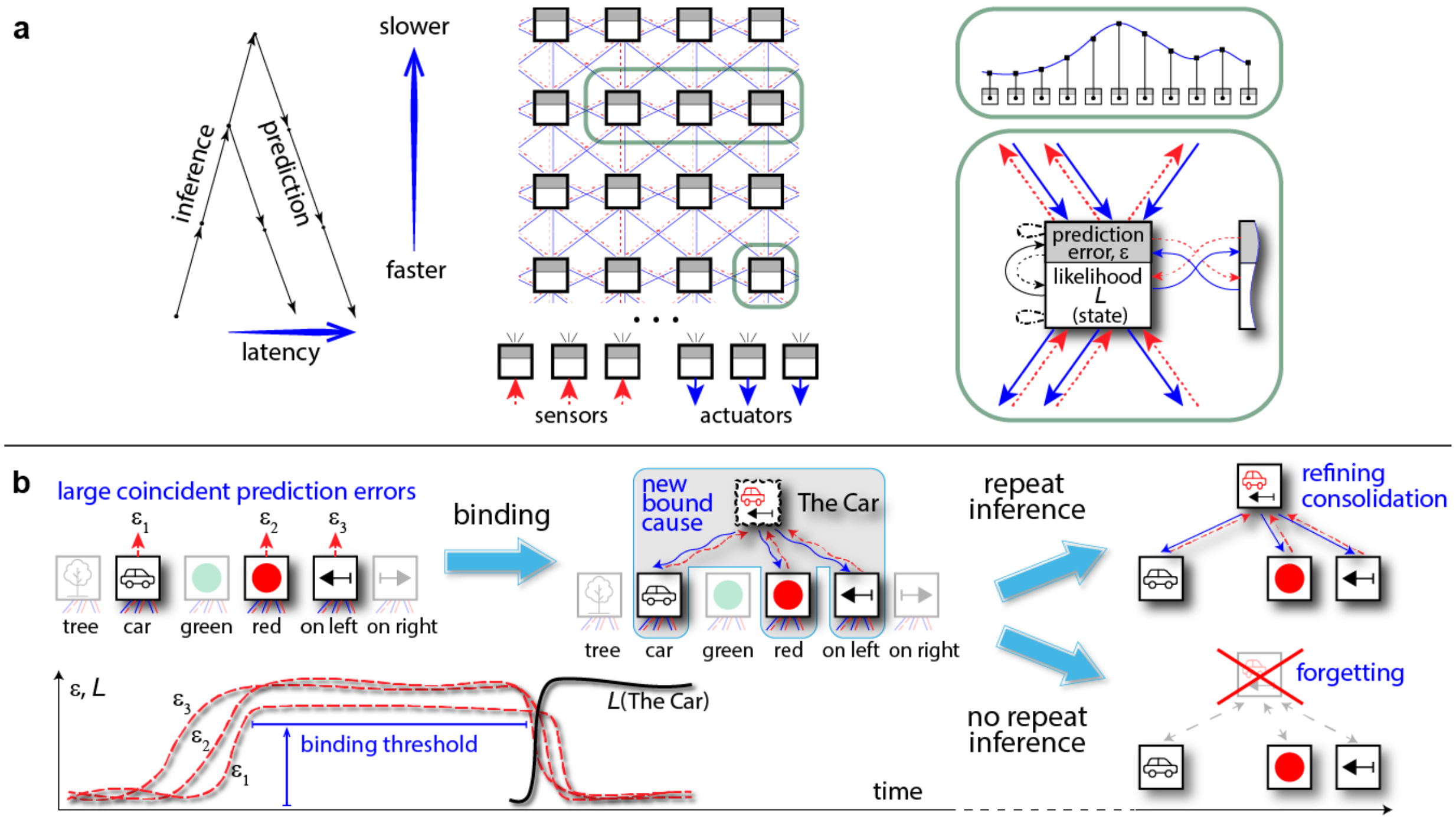

**Figure 1. Predictive processing and learning new causes by feature binding.** A generative predictive processing model (**a**) encodes the world and action state within a hierarchy of discrete connected cause units (middle). Each unit (right) encodes a scalar state of a single cause (e.g., a categorical cause likelihood) and the corresponding scalar prediction error. Multiple elements may interact to construct probability distribution maps (top right). The causes higher in the hierarchy predict the lower causes (blue solid arrows) and receive their prediction errors (red dashed arrows) for inference. The lowest levels represent sensory inputs with externally given states and actuation outputs with prediction-dictated states. To maintain predictive power in the presence of processing latencies (left), the cause dynamics at higher levels are slower than those at lower levels, such that higher levels encode more persistent and more abstract causes. (**b**) illustrates the theorized learning mechanism for new categorical causes. Whenever several feature causes, such as shape, color, or location, are inferred coincidentally and exceed a specific (prediction error) × (duration) threshold, such coincident unexplained causes are attributed to a new common cause by learning by binding. The common cause is provisionally added to the model, and immediately thereafter, the features that are bound are predicted. As time passes, if the new cause is inferred repeatedly, its predictions are refined and consolidated. Conversely, without repeat inference, the prediction strengths are gradually attenuated to zero, and the new cause is thereby discarded.

## Hierarchical learning by binding forms unified and differentiated access conscious contents.

As noted by(Rutar et al. 2022), conventional learning within PP is limited by the model's fixed causal structure, and additional structure-proper learning mechanisms are needed to enable human-like compositional and efficient learning. From the Bayesian parameter learning perspective, there is a scant rationale for large or discontinuous updates to parameters learned over multiple prior epochs, let alone discontinuous changes associated with near-single-example learning or the discovery of new causes. To overcome this limitation, we propose adding to PP a new type of learning for causal structure proper. Our learning-by-binding mechanism discontinuously adds previously unknown causes, that specifically remove the largest, most persistent and time-correlated prediction errors presently unreconciled within the model. The added causes are then retained or discarded depending on their subsequent usefulness. As illustrated in Figure 1b, when several significant prediction errors exceeding a binding threshold occur concurrently and persist for longer than a binding time interval, the model is modified by adding a new cause with prediction weights that precisely account for the specific prediction errors. This expresses an assumption—a learning bias—that causes with significant concurrent errors may be mutually related as features of a common hidden cause to be learned.

We may view this learning-by-feature-binding as a Bayesian update of the beliefs about the causal structure of the world. As a prior, the concurrent errors are coincidental, and we should retain the new cause only if additional evidence for it is subsequently observed: the new cause must be inferred repeatedly, thereby providing a consistent reduction of the model's prediction errors. Therefore, we propose that without repeat inference, the new cause's prediction weights are uniformly decreased with passage of time, effectively discarding the cause on a suitable short-term memory timescale. This Bayesian model reduction increases predictive power by minimizing the model's complexity. On the other hand, with each subsequent inference of the new cause, the prediction weights are updated following the PP weight learning rules. Initially, the weights are plastic and subject to large updates with each inference. As the evidence for the new cause accumulates, both the update size and the temporal attenuation rate decrease until only the small residual plasticity and zero attenuation are reached. Thus, the cause is consolidated, becoming a permanent addition to the PP model. One way to implement this within a locally connected hierarchical network is to have a reserve of latent causes that initially do not make predictions but are broadly connected and can be inferred in response to causes with large, concurrent and persistent errors within the level below. The latent causes may implement leaky integrators for these errors, and if a threshold is exceeded, they may abruptly adjust the prediction weights to start predicting these causes.

Our concept builds upon a similar type of structure learning that has been recently numerically studied(Smith et al. 2020) using an active inference model. New concept acquisition through model expansion, Bayesian model reduction, and improved generalization were shown, demonstrating the feasibility of this approach. While having important similarities, such as recruiting from a pool of latent causes, the studied model did not explicitly include temporal dynamics, and the time-dependent attenuation proposed here could not be considered for Bayesian model reduction. This attenuation offers an intriguing direct connection to the fading of the short-term memory contents.

Our causal structure learning is compositional – new causes are composed of known causes as their features. Moreover, the learning mechanism is intrinsically biased to exclude weakly related features since only features with large, above-threshold, prediction errors are bound, and, correspondingly, large new weights are set to strongly predict these features and fully eliminate these errors. The mechanism can also be explicitly biased toward further limiting the number of features bound into a single new cause, imposing additional topological sparsity on the learned PP models. The sparsity bias facilitates learning simpler causes that are generalizable across different context and useful as features for further compositional learning.

The binding time interval must be long enough to allow local free energy minimization to occur; the PP model must have enough time to generate the best possible local predictions from the known causes before new causes are introduced. This timescale is generally longer at higher levels of the model, where the dynamics must be slower to retain predictive power in the presence of larger processing latencies from sensory input and latency differences between different sensory modalities. Within longer binding time intervals, asynchronous unpredicted features are bound together into new causes that encode and predict the feature's relative delays within simple time sequences. We further discuss the timing and postdiction effects in conjunction with Figure 3.

Rather than forming one new cause at a time, we propose that the structure learning is both parallel and hierarchical. First, multiple groups of features are bound separately and concurrently into multiple new causes, based on temporal correlation of persistent prediction errors within each group. In Figure 2a, a 'red' 'loud' 'car' located 'on the left' is perceived as a new bound cause for the unpredicted concurrent perceptions of 'redness', 'loudness', 'car' and activation of a particular 'leftward' location on a spatial map. Perhaps head orientation and repeated visual saccades to this location, triggered by the auditory system's sound localization, create high temporal correlation of activation between a particular place 'on the left' in the body-centric-frame spatial map, the corresponding visual field map location, and the specific visual inferences of 'red' and 'car'. Concurrent auditory inferences describe the specific sound – 'loud', etc. – with the auditory localization activating the same body-centric location. All these multimodal perceptions have high prediction errors—they are not predicted—and are highly temporally correlated with each other and with (directing attention to) a particular spatial location. Thus, the system binds a new cause, 'The Car', which predicts a 'red' 'loud' 'car' 'on the left', now represented as a unified compound object. Similarly, a 'green' 'tree', of a particular 'tall' height, may be simultaneously perceived at a specific location 'on the right' as a new bound cause 'The Tree'.

Furthermore, a just-formed new cause cannot be predicted as a feature of any already known cause; therefore, there is a large prediction error each time the new cause is strongly inferred. When it persists for a sufficient duration, it is also bound together with other near-concurrent unpredicted causes into a higher-level new cause. This hierarchical binding process unites multiple new and previously known causes together, resulting in one-shot learning of a unified shallow tree causal structure culminating in a single root cause, as illustrated in Figure 2a. The root cause couples together a large fraction of the recognized yet unpredicted features of the present perceptual moment, encoding its specific differentiated generative description. This encoding leverages the already known PP causes as the perceptual code vocabulary. The depth and temporal duration of this sentence-like tree representing the unified present moment is determined by the PP model's a priori fixed maximum structural depth and the temporal persistence of the root cause at the deepest layer set by its binding time.

This work's central thesis is that consciousness is entailed by this compositional learning process, and the unpredicted inferences bound into the new shallow-tree hierarchical causal structures constitute access consciousness contents at any given moment in time. These structures are unified yet differentiated, which is a key hallmark of the content of consciousness(Bayne 2012). Once they are learned by binding within the PP model, these contents are associatively recallable (Figure 2b), maintainable in working memory (Figure 2c) and globally available (Figure 2d) for modulating subsequent PP inferences and active-inference actions because they establish strong new generative connections across disparate and previously unrelated features. Consciousness is the computational function defined by this compositional learning architecture. Variations in global hyperparameters governing this computational function manifest as differences in the state of consciousness(Bayne, Hohwy, and Owen 2016) (Appendix 4). While the details of the conjectured learning architecture remain to be analyzed by rigorous machine learning theory and numerical experiments, in the remainder of this article we argue that this architecture provides a unifying explanation for a remarkably broad range of functional observations associated with consciousness and paradigm cases of empirical consciousness studies.

The tree unifies these features into a perceptual episode, represented by the root cause. Adding this causal structure to PP at least temporarily allows subsequent associative recall of these features by prediction (Figure 2b): whenever a large enough subset of the bound features is perceived again with sufficient likelihood, their newly bound cause is inferred, which, in turn, predicts the remaining bound features. These features acquire a nonzero estimated likelihood, further affecting inferences and predictions of related PP causes, manifesting as conscious (bound) and unconscious perceptions, priming and action triggering. A single branch or the whole tree may be reinferred, thereby allowing the system to recall a specific aspect or a full episode. If no further binding occurs, the recall is unconscious, i.e., is not itself recallable. However, if the recalled causes have prediction errors of sufficient size and duration to be again bound into a new episode, this occurrence in turn becomes conscious and available for later recall.

In the simplified example in Figure 2b, a new compound cause 'The Car', was formed by binding. Subsequently, a verbal prompt for 'red' increases the precision of 'red' via auditory perception, and the system infers the 'The Car' to decrease the prediction error for the unexpected 'red'. This inference is stronger within the common perceptual context, i.e., when many of the other features of the recent episode-root-cause containing 'The Car' are also present. The whole episode is also partially recalled, predicting 'The Car'. Recalling 'The Car' activates its features, such as the 'on left' location, and modulates subsequent inferences.

A shallow tree categorical cause structure with a common root appears generally consistent with the level of semantic complexity of a human declarative episodic memory recallable as a single unit. If a new cause bound from an initial example is repeatedly reinferred in different contexts, it may be generalized by conventional PP learning rules to a broader example class, converting it from episodic memory into generalized semantic knowledge. In our account, learning-by-binding directly entails associatively recallable declarative memory, thus providing a natural Bayesian structure learning rationale for declarative memory formation and recall, inherently within the PP framework. Separate accounts for declarative memory storage and retrieval mechanisms are no longer necessary.

The (access) conscious contents introspectively appear unified because that is how they are bound and subsequently recalled and reported. However, the full unity of binding is not a requirement. At the sensory levels, multiple new causes are continuously formed by the binding of various concurrent unpredicted features, but most do not persist long enough to be further bound with others into an episode. They are not bound with, and not recallable associatively from, the other contents of the episode but nevertheless can later modulate perception and behavior. Such local, non-unified binding explains the experimental observations of visual stimuli learning without attention(Meuwese et al. 2013), connecting the proposed model to the recurrent processing ToC. As expected from our model, in the control experiments the masking of the stimuli disrupted binding and prevented their subsequent use in a test task. We note that these new bound causes, formed without attention, could not include as one of their features the action of attending to their spatial location, color, or another attribute, which, as we discuss below, is necessary to maintain them in working memory (Figure 2c). This explains why additional learning via feedback was required in the test trials of Ref.(Meuwese et al. 2013) to connect these new causes to the appropriate response actions.

The described learning-by-binding uses the ongoing inferences of known causes as features to compose new causes. This permits learning new causes that perceptually represent aspects of one's own perceptual processes, i.e., meta-representations. For example, a cause may be inferred to distinguish an ongoing sensory experience of an object from the mere presence of the object in the world: if one briefly looks away, the object is still present but is no longer visually experienced. At a higher level of complexity, the system may also infer causes representing the experience as belonging to its 'self' – this is instrumental for representing and distinguishing experiences and beliefs of others, a key ingredient of a perceptual theory of mind. In our schematic example in Figure 2, the new-bound causes for 'The Car' and 'The Tree' are bound with the meta-representation causes for 'I have' 'ongoing experience' to form a top-level episodic cause that includes 'I experience/see The Car (a red loud car on the left) and The Tree (a tall green tree on the right).'

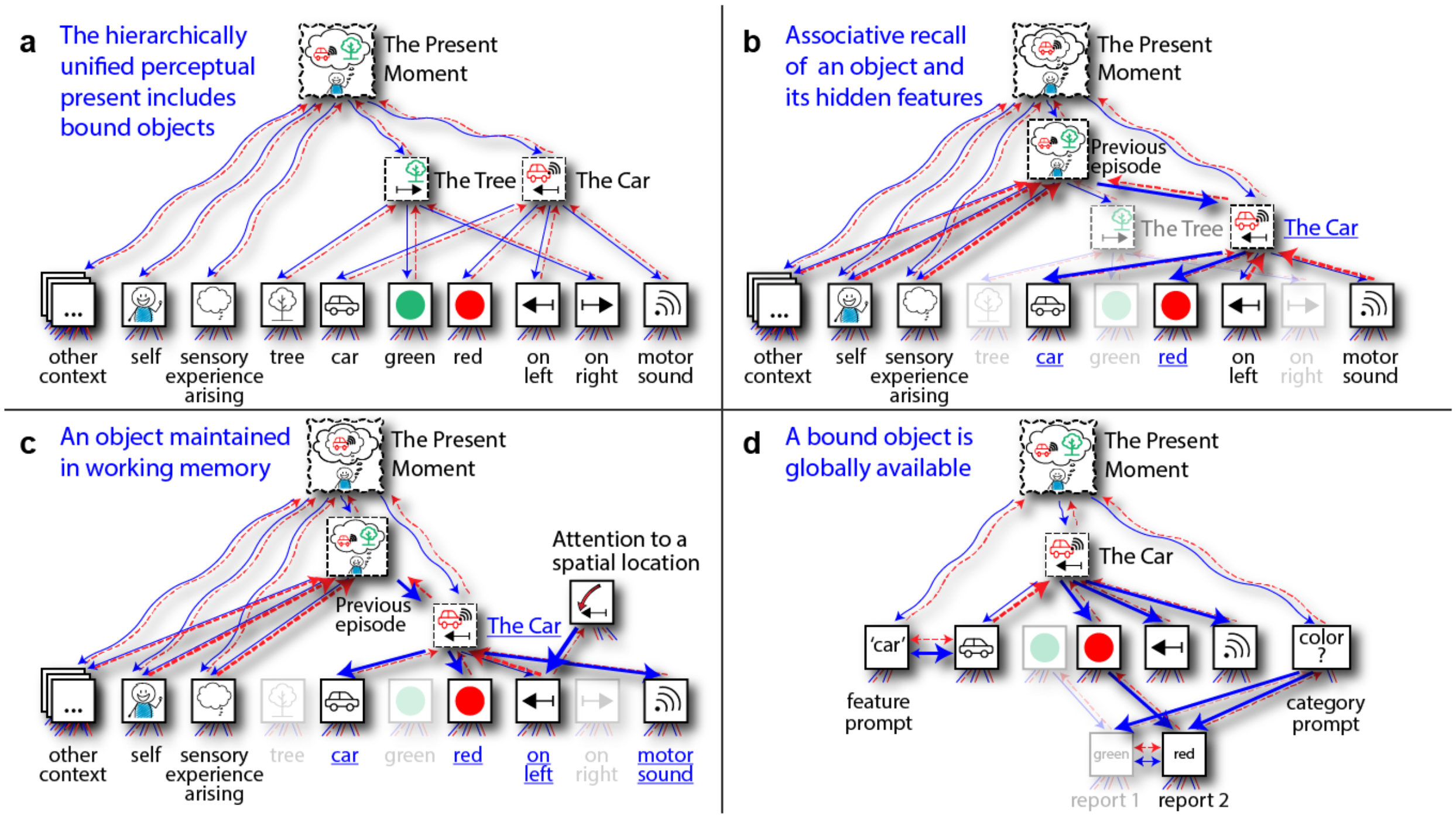

**Figure 2. Predictive processing combined with learning by binding entails commonly recognized mental phenomena.** (**a**) Learning by binding operates hierarchically, further binding multiple known and new object causes with large prediction errors into a single cause for the unified perceptual representation of the present moment—a shallow tree structure encoding the unified yet differentiated content describing an episode. (**b**) If some of the same features, such as the 'motor sound' localized 'on the left', are inferred again within a similar context, this may lead to the inference of the previous episode and 'The Car' object and the prediction of hidden features, such as the shape and color of the car, in the process of associative recall. (**c**) An action of attending to a feature, such as a specific spatial location, increases the likelihood of that feature. When one or more features of an object are thus predicted, the object may be maintained in perception even in the complete absence of direct inference from sensory data. This describes maintaining the object in working memory by selective attention. (**d**) When an object is formed by binding, its features predict each other. This causal connection between features makes the object globally available. In response to the prompt "What color is the car?", the feature prompt "car" increases the likelihood of the 'car' shape, which predicts 'red' via 'The Car'. 'Red' modulates the color reporting action triggered by the color query prompt.

## Quickly forming causal connections entails global availability.

The proposed learning by binding very quickly forms strong generative connections between previously unrelated causes (Figure 2d). Once bound, whenever the new cause is inferred from a subset of its features, the remaining, hidden features are generatively predicted. As discussed in the following example, this directly connects our theory to the global workspace theories of consciousness: the bound features temporarily become globally available by strongly cross-predicting each other and thereby immediately and strongly modulating many dependent perceptions and actions they had no previous causal connections to. The bound objects constituting conscious contents thereby become available globally.

In Figure 2d, after just a few hundred milliseconds needed to bind 'The Car', given a similar context, the 'The Car' may be inferred again from only the 'car' object, triggered by, e.g., a prompt for a 'car,' and predict the specific color, location and sound. Thus, the generic 'car' prompt starts to predict the 'red' color, 'left' location and so on, leading to their inference with nonzero likelihood. Within active inference, inferring 'red' modulates causally related actions such as reporting a color in response to a 'What color is the car?' prompt. Moreover, an alternate prompt, 'Point to a tree', might result in pointing to a specific location 'on right' bound to 'The Tree'. Thus, the bound objects comprising novel combinations of known features become available for a wide range of previously learned actions already connected to and modulated by each of these features. Thereby the new bound objects can immediately be acted upon in a large variety of complex ways. This type of flexible generalization is a notable advantage of the proposed learning architecture. It leverages only already known connections between action- and object-causes in the PP with active inference and requires neither information broadcasting distinct from the binding nor the existence of special neural codes.

## Working memory enabled by selective attention.

An important example of global availability is the ability to maintain a novel composite object in working memory. Within our view, this implies maintaining the object at high inferred likelihood, and thereby sufficient prediction error to be bound into multiple successive episodes. Active inference describes mental actions as changes in inferred likelihoods accomplished by top-down generative modulation. For a new object cause just added to the PP model this can be accomplished indirectly, by continuous active inference of the spatial location or another generic feature of the cause (Figure 2c). The object becomes available to the "attention spotlight" directed at its location or another bound attribute, where directing the spotlight to a specific location is an already known action.

The term "attention" is used to describe different processes in different contexts. In machine learning it typically is a multiplicative modulation in feed forward processing. In the PP and predictive coding with active inference it may describe either a generative (top-down) modulation of estimated precision, also having a multiplicative effect on inference, or a direct top-down prediction increasing the estimated likelihood for a categorical cause. For the following discussion, we use "selective attention" to refer to top-down active inference increasing the inferred likelihood of the attended cause, by either direct prediction or multiplicative amplification of weak evidence and noise. For a single categorical variable selective attention increases the estimated likelihood for the same available evidence. However, for a continuous variable, such as a spatial location, encoded as a distribution on a discretized grid (a spatial map) of categorical causes, selective attention both increases the likelihood of the attended location and lowers the likelihoods at other locations via the normalization relationships within the map, resulting in the increased precision.

In contrast to covert attention, overt attention to a location also triggers a cascade of active inferences that deploy the physical sensory apparatus to sample sensory data from that specific location. Such action might be triggered due to its high epistemic value, i.e., the high estimated salience of the location(Parr and Friston 2017). Overt attention results in simultaneous, strong and persistent inferences of multiple features perceived specifically from the attended location and may result in adding a new cause (e.g., 'The Car') to the model, which binds the location with these features. Continuous overt attention to the location keeps the likelihood of 'The Car' high, and its large prediction error leads to further binding of it into the episode.

Covert attention occurs without physical sensory modulation but nevertheless increases and maintains the new object's inference by sustaining a high likelihood for one or more of its bound features via active inference. In Figure 2c, attention to a specific location 'on left' results in maintaining a high likelihood for 'The Car', even without direct sensory evidence. 'The Car', having a large prediction error, is bound again as a feature of one or more recallable episodes and is continuously globally available. It is thereby maintained in working memory by active inference of its location.

This explanation of working memory relies on binding followed by reciprocal activation between spatial neural maps, generally located in the "front" of the cortex, and perceptual objects and features represented in the "back" cortex, consistent with neuroimaging observations in working memory tasks. However, there are other modalities of working memory—for example, auditory buffering—that may sustain bound auditory sequence information through repeat active inference of the first feature in the sequence triggering the bound sequence replay. This is facilitated by the same principle of binding the sequence into a single new cause and inferring the cause through one of its features, resulting in the generative replay. The described working memory is based on the cross-prediction of bound features. Its functionality is distinct from the storage and consolidation of declarative memory—the ability to passively retain the new bound causes within the PP model over an extended time. According to our theory, memory storage deficits, such as arising from hippocampal damage, are distinct from and do not grossly impair the learning by binding mechanisms underlying working memory and the global availability of the bound contents.

Considering the term 'attention' more broadly, we note that within the proposed functional architecture the perceived contents become globally available and enter the associatively recallable memory record via binding in response to sustained and unpredicted (i.e., 'unexpected') increase in their inferred likelihood. Hence, everything in the record has 'received attention' in a broad sense of a having undergone a sustained increase in the estimated likelihood. This includes bottom-up attention via normal perceptual PP inference, the described top-down selective attention to features via active inference, or both. While this link between broadly defined cognitive attention and binding in out account sheds some light on the intuitions underlying the notion of attention and the attention schema theory of consciousness(Graziano et al. 2020), we do not equate consciousness with either attention or the schema for it.

## Paradigm cases of consciousness research: stimulus timing, contrast, masking and mandatory integration.

To be bound into a new cause, the features must have inferred likelihoods significantly above predictions, and the resulting errors must exceed certain combined thresholds in size and duration (Figure 3a,b), as in the leaky integration and threshold evidence accumulation model. The hierarchical binding of new causes further, e.g., into an episode, requires additional persistence time. Therefore, sensory and cognitive manipulations that decrease the size and persistence of prediction errors interfere with the binding of inferences into conscious contents. This explains the effects in the classic experimental paradigms where an unpredicted object's stimuli are presented for a short time followed by masking to control the duration of the object's inference within the PP model (Figure 3a) and where the presentation contrast is manipulated (Figure 3b) both of which modulate binding in our model.

Our model also explains the paradigm case of perceptual integration, where conscious perception is manipulated by presenting a set of features simultaneously or in a quick succession. Within the conventional PP, integration occurs when the features are a posteriori predicted or, more accurately, postdicted by inferring a known common cause (Figure 3c,d). The known cause decreases the feature errors and in our architecture the features do not have sufficient error durations to be individually bound into a new cause. In contrast, the inferred known common cause may remain unpredicted long enough to become bound into a recallable and reportable episode (Figure 3d, left). Therefore, even though the features are strongly inferred by the PP for an extended period, they are unconscious and not associatively recallable separately from their inferred known 'integrated' cause. While this cause becomes part of the reportable and actionable conscious contents, the individual features do not. Specifically, our theory predicts that if perceptual integration occurred, it is impossible to recall which specific combination of many possible features was present and has led to the recallable instance of their common cause. Our theory accounts for the postdiction experiments showing mandatory integration of time-sequenced features(Herzog, Drissi-Daoudi, and Doerig 2020), as well as the unconscious integration of features into conscious wholes more generally.

As illustrated in Figure 3d, learning by binding specifically explains the apparent 'discreteness of time' in postdiction, as experimentally observed in Ref.(Herzog, Drissi-Daoudi, and Doerig 2020; Drissi-Daoudi, Doerig, and Herzog 2019). When a feature remains inferred and unpredicted for a sufficiently long period, it is bound and recallable. However, when two features are postdicted by a common cause before the first feature can be bound, neither feature is bound, and only their common cause may be bound and recallable (Figure 3d, left). If a third feature is inferred after the common cause for the first two features is inferred (Figure 3d, right), the prediction errors of the first two features are already low, and only their common cause and the third feature can be bound into a recallable episode. Therefore, even though it is presented in quick succession after the second feature, the third feature does not undergo mandatory unconscious integration and instead becomes part of the recallable conscious memory record, in agreement with experimental observations.

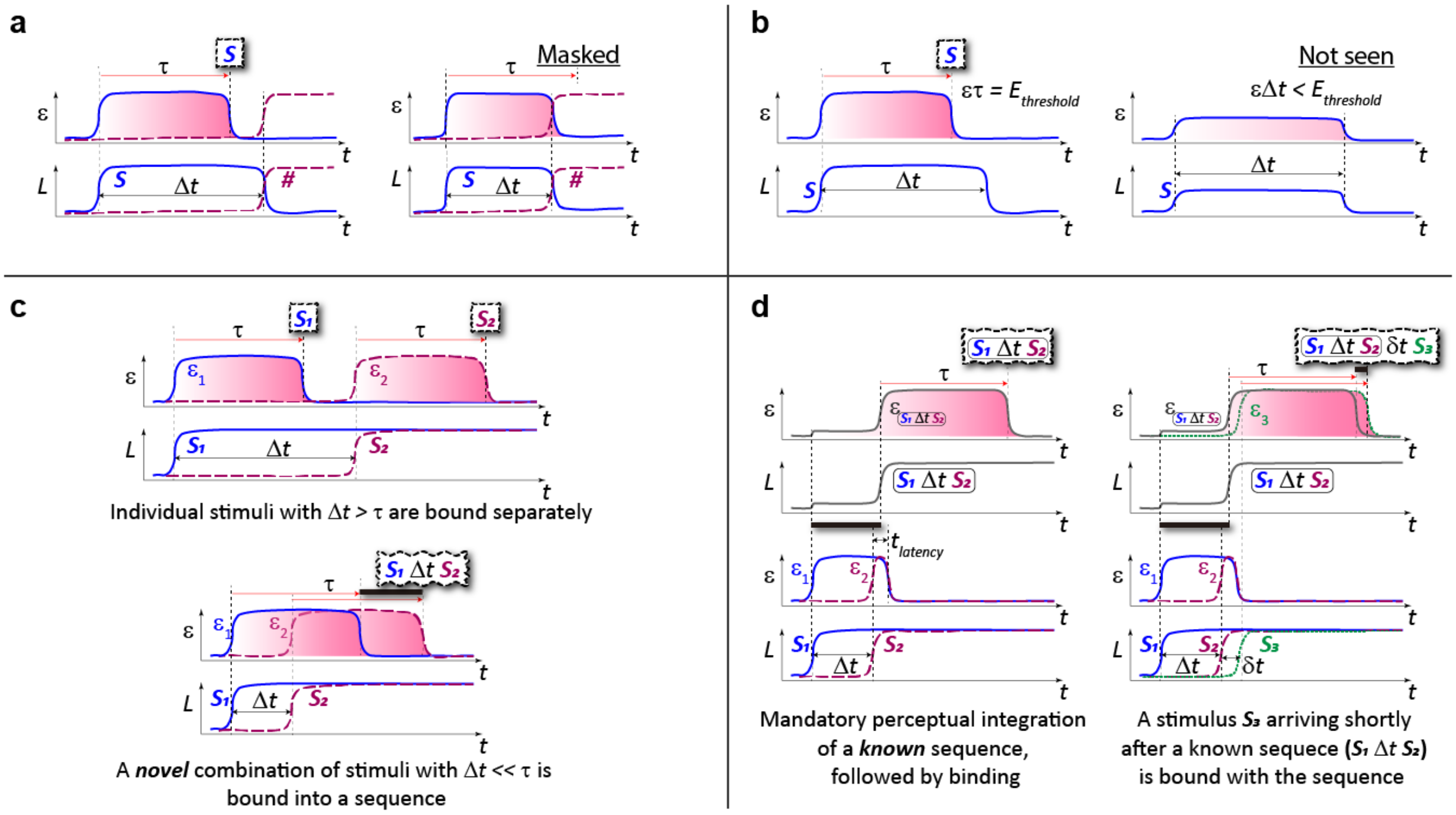

**Figure 3. Learning by binding explains time domain observations.** (**a**) For high-visibility stimuli, ***S*** binding (dashed box) occurs after a large prediction error $\varepsilon$ persists for a minimum time $\tau$, and masking introduced at $\Delta t > \tau$ does not interfere with binding (left). Masking at time $\Delta t < \tau$ prevents binding by decreasing the inferred likelihood of ***S*** (right) before binding occurs. The color indicates the leaky integrator output. (**b**) For stimuli with larger prediction errors, binding occurs faster (left). When contrast is decreased, binding is delayed and may not occur within the finite stimulus duration (right). (**c**) When unpredicted stimuli $\boldsymbol{S_1}$ and $\boldsymbol{S_2}$ are separated by $\Delta t > \tau$, they are bound separately (upper). When the separation $\Delta t$ is short compared to $\tau$, the near-coincident stimuli are bound together into a cause representing a sequence ($\boldsymbol{S_1}$ $\Delta t$ $\boldsymbol{S_2}$) with a specific separation $\Delta t$ (lower). (**d**) Left: if a sequence cause has already been learned, it is inferred and postdicts $\boldsymbol{S_1}$ and $\boldsymbol{S_2}$, reducing their prediction errors with a short latency time delay $t_{\text{latency}} << \tau$. This perceptual integration is mandatory. The sequence cause ($\boldsymbol{S_1}$ $\Delta t$ $\boldsymbol{S_2}$) is subject to further binding, while the individual stimuli $\boldsymbol{S_1}$ and $\boldsymbol{S_2}$ are not. Right: if a third stimulus $\boldsymbol{S_3}$ occurs a short time $\delta t$ after $\boldsymbol{S_1}$ and $\boldsymbol{S_2}$ such that $t_{\text{latency}} < \delta t < \Delta t$, the ($\boldsymbol{S_1}$ $\Delta t$ $\boldsymbol{S_2}$) will be inferred before $\boldsymbol{S_3}$, and after time $\tau$, they may be bound together into a new sequence (($\boldsymbol{S_1}$ $\Delta t$ $\boldsymbol{S_2}$) $\delta t$ $\boldsymbol{S_3}$). The mandatory nature of the ($\boldsymbol{S_1}$ $\Delta t$ $\boldsymbol{S_2}$) integration and the condition $\Delta t < \tau$ explain the experimental observation of the apparent discreteness of conscious perception over time.

Depending on the level in the PP hierarchy, the binding timescale may be between ≈200 ms (e.g., for low-level visual perception) all the way to, perhaps, ≈1 s for multimodal and amodal perceptions integrating higher-level stimuli with larger latencies. After that, the duration of the continuous or repeated inferences of these newly bound causes can be much longer, governed by the strength of sensory inference, the action of attention and the competition with perceptual distractors—the nonlocal dynamics within the PP model. The same new cause can thus be bound with multiple different other causes into a series of individual episodes. This longer timescale of attention-mediated perceptual dynamics spans at least ≈3 s and can likely be much longer with attention training(Srinivasan, Tripathi, and Singhal 2020). Thus, our model provides a mechanistic functional account of the three time ranges corresponding to the cinematic, extensional and retentional levels in Ref.(Singhal and Srinivasan 2021) by providing two distinct time scales—binding and attentional dynamics—to separate the time ranges.

## Conscious contents: scope, richness and phenomenality.

Is there any flexibility in defining which inferences should be called conscious contents (CC)? In our architecture, the functional status of all causes at a given time is fully determined by whether they are being (a) inferred with a nonzero likelihood and (b) bound into new causes. The vast majority of PP causes have a near zero likelihood and are not part of the CC. The known causes that are inferred but fully predicted by other known inferred PP causes only contribute to perception and action but do not participate in learning by binding and are not available globally. While contributing to priming, they should be defined as unconscious. Highly complex stereotyped unconscious PP perception and action can occur entirely without binding. Unconscious nondeclarative perception and action learning can also proceed via prediction weight updates following conventional PP learning rules. Such nondeclarative learning is gradual, in contrast to single-example learning by binding, which quickly and strongly connects unrelated causes.

This argument limits the CC to only the causes that participate in binding. A subset of such causes is hierarchically bound into a unified shallow tree structure representing an episode, making them globally available, satisfying a common criterion for access-conscious contents. Meanwhile, multiple new causes are continuously formed by binding and may or may not be further bound into the new episode, depending on their persistence. The formation of new bound causes that are never bound into an episode and do not become globally available reconciles the divergent views on the CC between recurrent processing and global workspace ToCs. Namely, such bound causes may be considered CC within recurrent processing but not within the global workspace ToC. Our functional description makes this distinction a matter of CC definition preference. From the perspective of discontinuous new-structure learning as the defining function of consciousness, the broader CC definition consistent with recurrent processing is operationally useful. However, the narrower CC definition consistent with the global workspace is also logically coherent and better fits the reportable, recallable and perceptually unified contents.

Given either definition, the question of richness or sparseness of CC becomes an empirically testable question of what contents are bound locally and globally under specific experimental conditions. In Sperling's experimental paradigm,(Sperling 1960; Sligte, Scholte, and Lamme 2009) multiple unpredicted visual symbols are all perceived (inferred), but only a narrow subset ends up being bound to and then actively maintained in working memory by selective attention to the experimentally prompted spatial location. While other symbols are not directly recallable, it is an empirically answerable question whether symbols outside the prompted subset have been bound to each other and/or to the concurrent unpredicted context. This question may be directly addressed using the learning without attention experimental paradigm(Meuwese et al. 2013) to determine whether these unprompted symbol combinations were learned and can later be recruited to perform a task. While the "unified globally available" CC definition excludes the larger symbol set, within the broader CC definition of "any bound inferences", the unprompted symbols may be experimentally determined to have been conscious by showing that the presented symbol combination can be subsequently distinguished or used to modulate behavior, as in Ref.(Meuwese et al. 2013). Thus, the extent to which learning by binding occurs without full unification and global availability can be experimentally determined, while whether to call such contents conscious is a matter of selecting the CC definition.

Another question is whether a category of 'phenomenal consciousness' contents can be defined or described within our functional theory. Specifically, is there a coherent definition of 'phenomenality' ascribed to the perceptual contents within our theory that reasonably corresponds to the relevant intuitions and is operationally useful for describing empirical data? One possibility is to ascribe phenomenality to the contents whenever they are bound together with particular types of inferred causes representing various aspects of the internal process of perception and cognition. These metacognitive representation causes are inferred, for example, to differentiate immediate sensory perceptions from perceptual facts without immediate sensory evidence and from endogenously generated, e.g., imagined, contents. Other bound meta-representation labels may differentiate perceptual inferences as belonging to the 'self' or 'another'. Such meta representations of 'sensory experience is arising' or 'I am having an experience' are learned and inferred to describe and predict internal perceptual contingencies but otherwise have the same standing within the model and are typically as transparent as perceptual representations for 'objects out there', learned and inferred to describe and predict regularities in sensory data and internal representations (Appendix 1).

For example, in the same way as 'red' is bound with 'car' to form 'the red car', the 'sensory experience of' can be bound with 'the red car' to form 'the sensory experience of the red car', a compound cause that includes 'phenomenality' as a feature. By having this feature, 'the red car' is represented as being phenomenally experienced. This type of phenomenality is just another perception, an inference the system makes to generate predictions. Unlike higher-order theories, we do not describe any such metacognitive inferences as granting (access) conscious status to other perceptions. However, the transparent nature of these metacognitive inferences explains the intuition that there is something that is undoubtedly, undeniably perceived (and therefore must be "objectively" theoretically explained), which is beyond the mere perception of objects and therefore appears beyond any functional theory of consciousness. Explaining the reason for this intuition is the meta-problem of consciousness(Chalmers 2018). It is solved by recognizing that the perception of phenomenality is a transparent perceptual inference, and to the perceiving system phenomenality appears no less real than its perceptions of outside objects, while resisting conventional 'objective' explanation (Appendix 2 and Figure A1).

## ACTION

How could a perception- and action-learning system be organized for online learning of deep hierarchical action policies maximizing gene proliferation in biological organisms by maintaining homeostasis, maximizing reproduction, offspring survival, group cooperation and other gene proliferating strategies? Combining reinforcement learning(Sutton and Barto 1998) with deep learning is arguably the most successful known approach for complex action optimization, with examples ranging from self-driving cars(Thrun et al. 2006; Kiran et al. 2022) to the best self-learning game-playing algorithms(Silver et al. 2018) to large language model chatbots. There is significant neurobiological evidence of reinforcement learning reward signaling in the brain(Montague, Dayan, and Sejnowski 1996; Schultz, Dayan, and Montague 1997; Redish 2004; Cohen et al. 2012), with recent studies(Jeong et al. 2022) pointing away from the overly simple time difference reinforcement learning (TDRL) of perceptions and cue-reward associations toward learning contingencies between past perceptions and "meaningful" events associated with significant and unpredicted positive or negative rewards. Finally, an RL framework provides a natural way to account for the subjective observation of perceptual valence.

PP models can naturally encode perception-action policies via active inference(Parr and Friston 2019; K. J. Friston et al. 2010; K. Friston et al. 2017). The relationship between active inference and RL is an open area of research(K. J. Friston, Daunizeau, and Kiebel 2009; Tschantz et al. 2020; Millidge et al. 2020) with compelling arguments that active inference subsumes RL as a special case, while providing a principled approach to the exploration-exploitation tradeoff. As a special case, TDRL explicitly computes the total reward prediction error and its global signaling may enable more efficient online learning in large models. Within this context, we suggest several logical principles for how a system combining PP and TDRL might be organized. Within explicit RL, accurate and unbiased estimation of future value is key for learning actions that maximize the reward. Thus, even if action and perception are encoded within a common hierarchy, their functional roles are separate. Perception is learned to accurately estimate the expected on-policy future value and to represent the regularities of the world that are helpful for controlling action. Meanwhile, action policy is learned to maximize future value without undue decrease and preferably with an increase in perception accuracy via maximizing the epistemic value.

In our PP-RL model in Figure 4, the PP system with learning by binding, as introduced above, encodes deep perception-action policies. We assume additional, separate, and comparably simpler circuits detecting survival- and reproduction-relevant events and signaling immediate positive or negative rewards of several different kinds. To implement RL, we first segregate the PP model into two areas with distinct learning rules for perception and for action, with rough qualitative similarities to the “back” and “front” cortices. Perception is learned for the dual purposes of guiding action and estimating the expected cumulative reward (future value) for the TDRL (equivalently, expected free energy). In contrast, action policy learning uses a perceptual future value estimate to maximize the future value. To account for the multiple kinds of emotional/perceptual valences, we suggest that multiple kinds of future values are estimated simultaneously and separately and then summed to obtain a scalar positive or negative total value surprise signal, which is broadcast to the action-encoding cortex for reinforcement learning. Putative corresponding brain circuits include the amygdala connected to the perceptual areas for value encoding and estimation and the value system nucleus accumbens and ventral tegmental area for dopaminergic signaling of value surprise reinforcement mostly to the front of the cortex.

While detailed analysis of the PP-RL architecture is beyond the scope of this work, here we note the two basic learning loops in Figure 4. The first broadcasts the separate reward prediction errors for each reward kind to the perceptual area of the PP model, where each recently inferred cause can use it to modify its own value prediction, locally following TDRL value learning. Notably, the goal of perception is to accurately estimate and predict both positive and negative values and rewards. The second loop adds the errors from different value kinds and broadcasts the sum to the action policy encoding causes, where the learning goal is to suppress any recent policy modifications that may have resulted in a negative value surprise and to reinforce policy modifications in response to a positive surprise. Given the enormity of the action space, selecting the optimal action in a given state via Q-learning(Sutton and Barto 1998) appears intractable. On-policy action learning must proceed by policy modification and testing for resulting value surprise. Intriguingly, learning by binding that binds action causes to perception causes may be seen as one such policy modification.

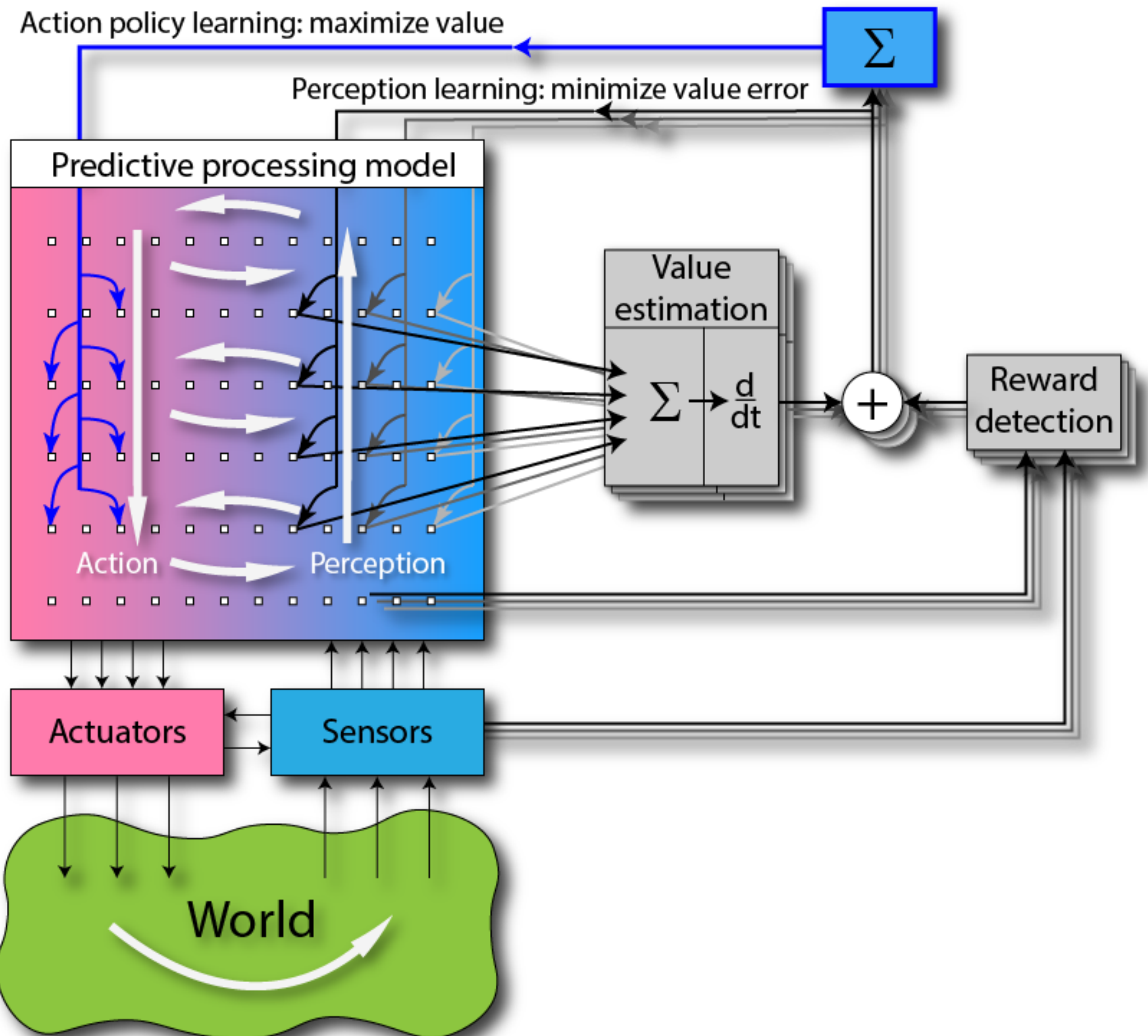

**Figure 4. Predictive processing with active inference learns to maximize the estimated cumulative future value via reinforcement learning.** The predictive processing model combines perception and action, which have distinct learning goals: perceptual causes predict future value accurately, while actions maximize it. Several types of immediate rewards are calculated directly from external and body sensory data. Each perceptual cause can contribute a positive or negative value estimate of each value type. Predictive value estimation via perception is learned by attributing reward prediction errors backward in time to recently active perceptual causes. Action policy is improved by increasing (decreasing) the future likelihood of advantageous (detrimental) actions: overall, positive (negative) reward prediction error is used to sensitize (inhibit) the connections that triggered the recently active actions.

## DISCUSSION AND OUTLOOK

We have argued that the major scientific and first-person observations giving rise to the concept of consciousness can be explained by considering a computational functional system combining generative perceptual modeling, such as predictive processing, with learning by binding and reinforcement learning. Our qualitative theoretical proposal comprehensively addresses major empirical phenomena of consciousness and meets hard criteria(Doerig, Schurger, and Herzog 2021) for a theory of consciousness (Appendix 3). It retains and unifies key insights from current leading theories of consciousness within a single and specific computational architecture. Our arguments already show that the proposed architecture possesses 10 out of 14 indicator properties of a computational conscious system distilled via a comprehensive review of current theories of consciousness(Butlin et al. 2023). Of the remaining four indicator properties, AE-1 and AE-2 are already partially addressed by the PP-RL architecture, and the indicators HOT-3 and AST-1 can, in principle, be met by our system at a sufficient level of complexity.

In this work, we make qualitative arguments about the properties entailed by the proposed learning system and how these properties explain relevant empirical observations. In these descriptions, we preferred to be specific and possibly wrong rather than vague and unfalsifiable, aiming to facilitate empirical testing. Undoubtedly, many of these descriptions must be refined and corrected, but we hope that the main ideas will withstand the empirical tests. Independent of these details, the key result of this work is the roadmap for how a comprehensive computational functionalist theory of consciousness may be conceived.

The proposed architecture is amenable to rigorous mathematical description, numerical modeling, and formal theoretical analysis within the frameworks of statistical learning theory, machine learning and predictive processing. While much remains to be understood and refined, numerical testing of such computational functional architecture is, in principle, straightforward. However, it is critically important to give proper ethical considerations to such research, particularly as modern computational resources enable testing of numerical models with a very high level of complexity. The ethical goal of gaining knowledge and reducing epistemic indeterminacy should be weighted carefully against increasing the risk of creating unnecessary artificial suffering(Metzinger 2021). Theoretical analysis of this proposal may shed new light on the nature of conscious suffering, providing the necessary guardrails for future research. This work brings closer the goal of full scientific understanding of human consciousness, with potential for transformative philosophical, cultural, societal, and technological implications.

## ACKNOWLEDGMENTS

I am grateful to Prof. Paula Droege for many insightful comments and suggestions on the earlier version of the manuscript. I would like to thank Dr. Matthew Daniels and Dr. Ishan Singhal for their comments and suggestions and Mr. Jim Antonisse for many hours of stimulating conversations on consciousness, machine learning, perception, modeling and beyond. I would like to thank Prof. Gregoire Sergeant-Perthuis for insightful questions and Prof. Karl Friston for extensive and valuable comments following my earlier presentation of this work, and for suggesting additional useful references.

# SUPPLEMENTARY INFORMATION

## APPENDIX 1. Transparency of perceptual models, higher-order theories, 'experience' and 'self'.

The proposed functional architecture learns ever more sophisticated Bayesian generative models of the time-dependent sensory data stream. This model construction process is hierarchical, adding new causes describing regularities among the known causes. The inferences of the current model serve as inputs for further model construction in a process that is, in principle, limitless, while in specific organisms, the model depth and structure are limited by the particular neural architectures and processing latencies of their brain.

Learning by binding creates an associatively recallable record consisting of the variously bound causes that were not fully predicted at specific times. As a matter of this perceptual record, these bound causes are the only things that compose it and can be recalled by the system. In most cases, these perceived and recallable causes are not additionally recognized and metacognitively represented as outputs of its perceptual models; rather, they appear to the system only as elementary units constituting the perceptual record. In other words, the models giving rise to these inferences are both transparent(Metzinger 2003) (not directly perceivable) and cognitively impenetrable by the system, for as long as the system is unaware of its own functional organization and does not have a suitable perceptual model for recognizing and representing its own functioning beyond the causes classifying types of experiences and distinguishing experiences of self from experiences of others within a basic theory of mind.

As with all causes, the perceptual inferences discriminating 'sensory experience,' 'indirectly inferred experience' and 'imagined experience', together with the inferences about other specific unpredicted characteristics of perception at each moment, are subject to being bound into the perceptual record by the learning process. Thus, the perceptual record includes these directly perceived meta representations of the perceptual process. Somewhat similar representations are discussed within higher-order theories of consciousness(Cleeremans et al. 2020; Cleeremans 2011; Brown, Lau, and LeDoux 2019). However, rather than postulating them, here, we argue that these types of meta perceptions are entailed by our learning-by-binding functional architecture as inferences within the learned hierarchical PP model. In contrast to higher-order theories, these representations do not determine whether a given low-order perception is conscious, such as entering the declarative memory record or modulating a large range of actions via the global workspace-like effect of binding. In our view, systems with simple models having limited or no metacognitive representations can have the binding and the resulting global availability and associative recall functions we associated with consciousness. However, the metacognitive inference of 'sensory experience arising' amounts to the system representing its low-order perceptions as having a 'phenomenal' character, i.e., as being 'experienced'. Notably, the higher-order model only recognizes a feature of the lower-order model output and does not represent that this output is produced by modeling, thereby allowing the lower-order model to remain transparent.

Transparent perceptual models span a range from basic sensory processing, such as illumination-invariant color recognition, to highly complex abstract models, such as the model for 'self' and the related "phenomenal models of the intentionality relation"(Metzinger 2003; 2005; 2020b). A broad range of self-models, including procedural and declarative in addition to the perceptual ones, have been discussed as implicit beliefs comprising m-consciousness(Graziano et al. 2020). Following this line of reasoning, a sufficiently sophisticated system may learn to infer theory-of-mind ‘agent’ causes for describing people and animals; furthermore, learning to perceive itself is one such agent. Thereby it learns to infer the cause of ‘self’ as an ‘owner of experience’ (“I have an experience”) and the ‘source’ of action and infer ‘beliefs’, ‘motivations’ and other properties ascribed to such agents and to the ‘self’. Based on these perceptions, the system may also learn the actions of reporting and discursive reasoning with the belief of having/being a ‘self’. This view of the self parallels that described by Thomas Metzinger(Metzinger 2003), Tim Bayne(Bayne 2012) and others(Graziano et al. 2020). As a cause inferred within a transparent model, the ‘self’ is clearly and undeniably perceived as truly existing, as a matter of systems’ perceptual record. However, this perceived ‘self’, like many naïve perceptions, is an illusion in the sense that reality is different from its representation. Prior to an accurate scientific theory of information processing in the brain, which is the ultimate goal motivating this work, ‘experience arising’ and ‘self’ are naïve and transparent perceptual models learned by the system as part of perceptual modeling describing its own functioning.

## APPENDIX 2. The meta-problem of consciousness.

The meta-problem of consciousness(Chalmers 2018) is the problem of explaining why we may think there is a hard problem(Chalmers 1995) of consciousness. To solve the meta-problem is to explain why, within a particular learning system, the *perception* of 'experience' necessarily arises and why this perception appears to resist reductive explanation. Within our theory, the model making the higher-order inference 'an experience is arising' is a transparent model of the same kind as sensory perception models inferring 'trees' and 'cars' (Figure A1). For example, the system simply represents 'sensory experience' as presently arising with high likelihood rather than recognizing this inference as resulting from any modeling or reasoning. The recognition of 'an experience arising' is of the same kind as the recognition of a 'tree' or a 'car', as a matter of its functional consequences within the system, such as the global availability or the perceptual memory record. Whatever the system learns to recognize, whether about the outside world or about itself, appears to the system as experientially given truth, at least until it can learn to perceive and represent otherwise. As the system makes inferences about its own processes of perception and learning, 'experiencing' appears (to it) to undeniably exist in exactly the way 'cars' and 'trees' appear to exist.

We suggest that the 'explanatory gap' difficulty lies in the inability of previous theories to envision a mechanistic functional description entailing the occurrence of direct metacognitive perceptions of the 'experience arising' and 'I am having an experience' of the same kind as perceptions of objects and events attributed by us to the outside world, such as 'cars' and 'trees.' Meanwhile, from our internal perspective, both 'trees' and 'experiencing' are equally apparent perceptions. The metacognitive perceptions appear to us as directly perceived, the same way physical objects are directly perceived, therefore demanding the same kind of 'objective' explanation as used for physical objects. Most people outside philosophy do not routinely question the 'objective' existence of cars and trees, and even fewer perceptually recognize them as mere constructs—preferences from learned phenomenological models constructed by our minds to predict raw sensory data. Similarly, people feel the 'experience arising' as no less than an 'objective' fact. Physical theories describe external objects as being 'out there', independent of the process of perception. Thus, we expect and demand the same kind of objective explanation for the perception of 'experience arising,' yet without referring to the process of perception, this is logically impossible.

Here, we provide the same kind of description for the object perceptions and the metacognitive perceptions, both of which are Bayesian inferences made by the system to predict sensory data (and are distinct from physical-science objects as predictive descriptors within formal scientific models for the various aspects of reality). It has been pointed out that one fallacy leading to the appearance of the hard problem illusion is the treatment of the perceptions of physical objects and sensory events inferred with high certainty as direct, veridical representations of physical reality rather than as constructs we learn to infer for their predictive power(A. Clark, Friston, and Wilkinson 2019). These constructs are based on the outside world only insofar as predictable regularities in the sensorium-dependent sensory data are described using a particular hierarchical causal modeling structure—the hypothesis space of the learning system implemented by the brain. However, rather than attributing the meta-problem puzzlement to the role these especially certain mid-level representations play in the PP inference process(A. Clark, Friston, and Wilkinson 2019), we emphasize that they are additionally bound with the equally certain representations of 'experience arising'. We suggest that these additional perceptions, inferred but not recognized, give raise to the intuition of there being an explanatory gap, which is filled by accounting for them – recognizing the direct, transparent perception of there being an experience. This perception of ‘experience’ is simply another learned perceptual construct inferred to describe predictable contingencies in the perceptual modeling inferences at a lower level. The functional, mechanistic account of the process of perception and learning proposed here explains how the directly felt ‘I am experiencing’ becomes part of the system’s perceptual record and reportable working memory. Inferences made by transparent perceptual models appear to the system as objectively existing things and events, including the ‘experience arising,’ the ‘self’ and the ‘consciousness’ in the ‘I am experiencing’ and ‘I am conscious’.

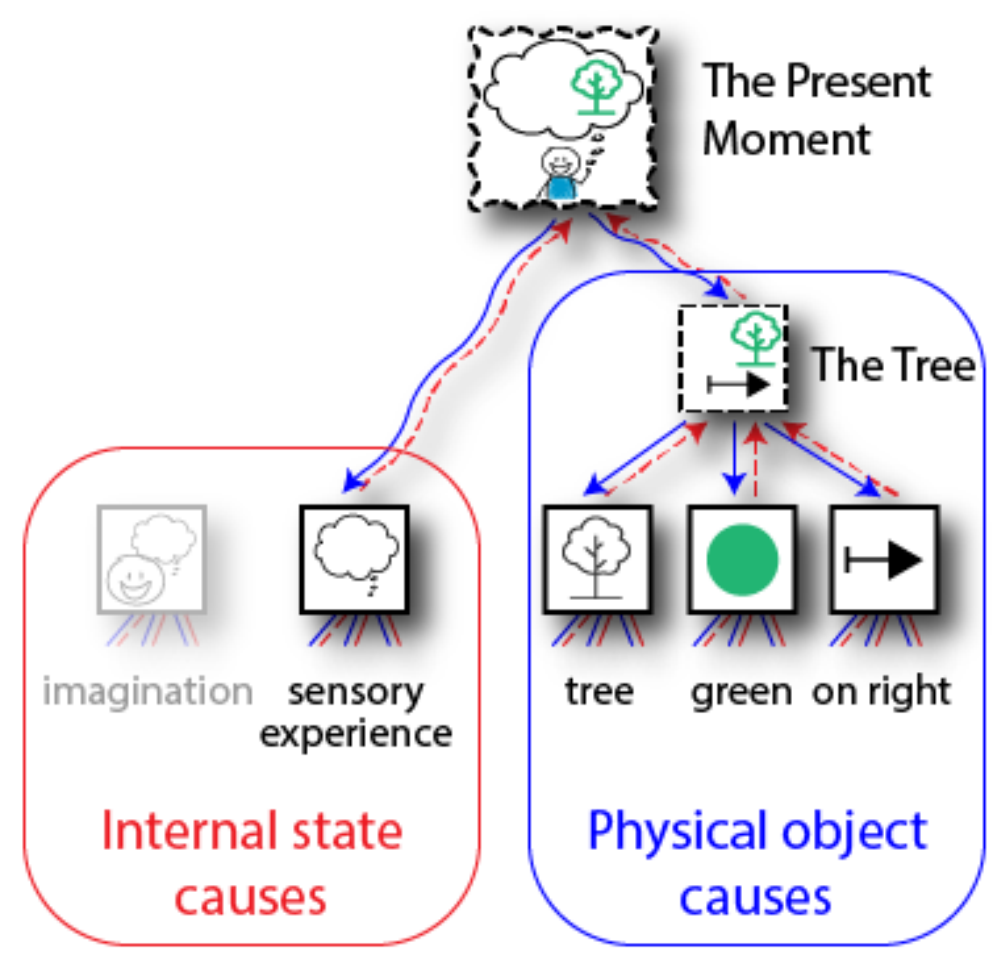

- Inferences, arising from lower level causes
- Appear as independently existing (transparent)
- Cannot be unperceived

Demand the <u>same type of</u> description yet,

| private “subjective” | shared “objective” |
|---|---|

Solution:

Causes are learned and transparently perceived by the system interacting with the world.

The system’s functional organization <u>entails these perceptions,</u> both internal and external.

**Figure A1. Meta-problem of consciousness.** The system makes perceptual inferences about the external world and about its own internal states, such as to represent whether an object is sensorily experienced or imagined. The internal state cause for “having a sensory experience” has the same functional properties as the physical object cause for “The Tree”: both are clearly and undeniably apparent and cannot be unperceived. Without understanding the nature of perceptions as learned and inferred causes describing internal and external states, the internal and external perceptions appear qualitatively different. For example, physical objects are understood by naïve observers as properties of the outside world independent of the system perceiving it. Importantly, a sufficiently complex system with this functional

organization must learn to perceive internal states and thereby be able to perceive, remember and recall having experiences.

# APPENDIX 3. Meeting the Hard Criteria for a theory of consciousness

Here, we describe our proposal according to the hard criteria for a theory of consciousness(Doerig, Schurger, and Herzog 2021).

## Empirical phenomena of consciousness being addressed by this work.

1. Our theory addresses both the content and the state of consciousness by describing (1) the functional mechanisms necessary and sufficient for consciousness (state) and (2) the formation of perceptual content and which perceptual content is unified, enters the memory record and is available for action, and which is not (content).
2. The conscious state is governed by hyperparameters of the various learning functions (the prediction error size, time persistence and correlation necessary for binding; parameters governing the ongoing inference of PP; the bound cause forgetting and consolidation rates; the value learning and action learning timescales and rates; and the exploration vs. exploitation tradeoff). All these global parameters modulate the state of consciousness continuously, and their different values distinguish conscious states.
3. Consciousness is unifying in the sense of (temporarily) attributing various contents to common causes and thereby (temporarily) constraining these groups of perceptions to covary. This unification is hierarchical, and at the highest level the contents are unified into an episode. Given the proposed binding rules, binding between low-level perceptual features may occur without further binding with higher-level features and the episode, accounting for the experimental observations(Meuwese et al. 2013).
4. The theory is temporally continuous but posits the existence of a time threshold for binding. Discontinuously varying outcomes are predicted, which depend on whether the time periods between perceptual inferences are shorter or longer than the binding time threshold, accounting for experimental observations(Herzog, Drissi-Daoudi, and Doerig 2020).
5. Unconscious contents are not bound or unified; therefore, they can have causal influence only on other contents with which they have a prior learned association, e.g., priming and triggering of previously associated actions and perceptions.

## Meeting the Hard Criteria.

1. <u>Paradigm cases</u>: The theory addresses several experimental paradigm cases as described. It is comprehensive and experimentally falsifiable.
2. <u>Unfolding and structure vs. function</u>: Theory is functional and does not limit how the functions are implemented. However, consciousness is a property than cannot be meaningfully ascribed to any separate part of the full system's functional organization, which must include future value estimation and action learning as well as the PP and binding. To function, such a system must be connected to a world via sensors and actuators, and rewards must be provided.
3. <u>Network size</u>: The theory describes a functional organization that implements consciousness, irrespective of size. Accordingly, small networks implementing this functional organization are deemed conscious, even if this consciousness is simple and limited in what it can represent, learn or enact. Each instance of a complete implementation of this functional organization within a large network is deemed separately conscious. Other than split brain patients, there is no clear evidence of multiple instances of this organization within a single brain. It is speculated that two largely separate instances of this functional organization exist in split brain patients, thus containing two consciousnesses. However, when the split brains are interconnected, they can no longer be considered separate and independent implementations of the functional organization. A normally connected brain implements a single instance of the functional organization.

4. Other systems: Within our computational functionalism view, multiple implementations are possible, including nonbiological ones. An appropriately functionally organized numerical model must be considered conscious.

# APPENDIX 4. States of consciousness.

The hyperparameters of learning by binding—the minimum prediction error size, activation duration and maximum asynchrony between features that can still lead to binding—varied across sensory domains and PP hierarchy levels, such as between single-domain, multimodal and abstract (amodal) mental objects. In biological brains, they are optimized by evolution to efficiently infer typical structures in sensory data within each domain, particularly those structures that are most useful for guiding evolutionarily advantageous actions. These binding parameters, as well as the plasticity of prediction strengths vs. cumulative activation and consolidation/forgetting timescales, can be empirically elucidated, such as by analyzing data from existing cognitive studies. These parameters directly affect both the perception and learning of events over time. The hyperparameters are modulated on short and long timescales by global state variables, such as the affective states and exploration-exploitation balance in RL. The major functional role of perception is to estimate future value for reinforcement learning an optimized action policy encoded within the same PP generative model. Therefore, on short time scales, high positive or negative valance strongly and dynamically modulates the learning-by-binding parameters to provide both value-relevant perceptual learning and value-increasing action policy learning.

In our theory, the slow global variations of hyperparameters controlling prediction and inference, PP learning and learning by binding, modulated by the value system, define the state of consciousness(Bayne, Hohwy, and Owen 2016). While specifying the exact dependencies is well beyond the scope of the present work, one important concept is the exploration-exploitation tradeoff. Lowering the inference threshold for action may lead to off-policy action generation for exploration. Perceptual inference might be modulated similarly. Additionally, the duration, level of prediction error and its correlation leading to binding may also be modulated. Some versions of these variables may be signaled by the tonic activity of the brain's aminergic systems, such as the VTA dopaminergic signaling for actions and serotonin signaling for perception or binding, potentially shedding light on the hallucinogenic effects resulting from pathological or pharmacologically induced imbalances in such signaling.

Considering perceptual value estimation, it is important to recall that much of the perceptual inference is unconscious. In our view, unconscious perception remains valanced and contributes to the overall future value estimate. Some fraction of this unconscious perceptual content is stable in time, continuously or repeatedly inferred over long periods. When these unconscious percepts have large specific positive or negative values along one or more emotional valence space dimensions, they provide continuous input to the value system. Excessive chronic positive and, particularly, negative bias may affect the system's hyperparameters, including offsetting the exploration–exploitation balance. This is one way our model connects to mood and its disorders, such as depression. Notably, if such unconscious inference becomes conscious in a type of perceptual shift that allows one to continue to perceive a cause while no longer fully predicting it, the inference participates in learning by binding and both its perceptual inference, and its specific value may be rapidly changed by the single-example learning mechanism.

In focused attention and mindfulness meditation training, the assignment of a high specific value to a goal of continuously inferring the meditation object with a high level of likelihood gradually modifies the policy and the perceptual model to reduce the inference of distractors. Over time, this makes possible a large reduction or even elimination of most perceptual inferences arising either from sensory data or from actions of imagination, except for those representing the task-relevant state of the system and the goal(Laukkonen and Slagter 2021). In a generative PP model, this also means a reduction in sensory and motor predictions. Sensory maps may have regularizing normalization relations between causes, possibly reacting to such tonic reductions in predictions and inferences by decreasing inference thresholds and increasing background likelihoods, which may explain accounts of vivid internally generated perceptions at certain stages of meditation practice and the uniform visual illumination reported by experienced meditators. Furthermore, lack of perception equals lack of input for future value estimation. When the goal of effortless focused object perception without distractors is initially achieved, the accompanying positive evaluation of the task performance is the dominant positively valenced perception, providing a singular positive input to the value system. In the absence of any other concurrent inputs, this is consistent with the meditative rapture typically described upon reaching this stage of practice. With further practice, the estimated future value of being in this state decreases, eventually giving rise to affectively neutral equanimity. Hypothetically, the reduction in the unpredicted perceptual content in deep meditation would mean progressively simpler episode content, consistent with reports of “pure awareness,” ultimately resulting in minimal phenomenal experience.(Metzinger 2020a) Notably, in a fully predicted state with no inferred causes with high prediction errors, there would be no episodic binding and no subsequent associative recall, consistent with the cessation events reported in some contemplative traditions.

In somewhat related “flow” states(Csikszentmihalyi 1975) are the states where only comparably narrow perception and action subspaces are being occupied, while much of the ‘self’-referential default mode perception and action content is temporarily not being inferred to avoid task interference. The future value in a flow state is estimated only from this narrow task-relevant perceptual content, which is neutral or positive when the task is being successfully executed. Notably, the associated positive affective state may differ markedly from the affective state in the default mode.

Sleep is known to be important for off-line learning. Consistent with our hypothesis, there may be offline regimes that activate parts of the PP model and/or change the prediction weights in the absence of binding and therefore fully unconsciously. Action and object encoding parts of the PP hierarchy may be activated differently or not at all. In contrast, dreaming appears to be the result of action exploration in response to tonic dopamine signaling to the front of the brain, whereby in the absence of sensory input, the explored actions of imagination result in hallucinatory perceptions. This has been argued to constitute a Bayesian model reduction(Hobson and Friston, n.d.) from the active inference perspective. An intriguing particular case is that dreaming might serve as offline RL value-learning iterations. In TDRL, multiple iterations are necessary to propagate the value backward over large time delays to assign it specifically to one or more of the predictive causes. By repeatedly replaying perceptual models forward in time, such specific values of perceptual causes can be learned. Temporary binding during dreaming is likely necessary for correct dynamic replay, resulting in conscious perception of dreams, while declarative learning from dreams is inhibited by modifying the hyperparameters to disable memory retention and consolidation, inducing sleep amnesia.

# APPENDIX 5. Measurement of consciousness.

Generally, we propose that the defining consequence of consciousness is efficient declarative learning that is not accessible to unconscious processes or systems. Therefore, the empirically testable presence of these types of learning modalities in biological organisms can serve as a measure of consciousness, separate from and broader than introspective reports. Such measures taken together with other cognitive and neurophysiological data and numerical models provide a path for developing a validated theoretical framework for consciousness.

According to our theory, recallable conscious content and conscious action arise through the specific interactions of the PP, binding, future value estimation and RL; therefore, experiments might attempt to isolate and target these specific processes in the brain for both measurement and controlled manipulation. PP without binding is unconscious. At the lowest level, a conscious perception is a binding event that creates temporary cross-prediction between previously unrelated causes. A mere correlation is not sufficient to confirm the new causal connection; rather, one of the bound causes must be manipulated, and the effect on the other must be measured, such as in an associative recall. Importantly, studies should focus on binding between causes that were previously unrelated.

This binding may be studied both at the low level of the PP, the presumed result of recurrent processing in low sensory layers, and at the high level of the PP hierarchy, where the content is fully unified and globally available. Experiments may manipulate the inference of the causes that are being bound, such as specific perceptual features at the sensory PP level or specific actions of selective attention to attributes at the higher level. The manipulations can target the strength and duration of both perceptual and attention action inferences—stimulus contrast and masking for perception and distractors to trigger interfering attention actions for attention. Naturally, these are already common experimental paradigms, but our view highlights the need to ensure the novelty of the presented combination of stimuli, i.e., controlling for preexisting PP causes for the combination. We advocate measurements of binding of specific, controlled features to each other rather than measurements of the less- controllable binding of a feature to the whole experimental context, which is often the case in the present paradigms.

When considering measures of the conscious state, such as the perturbation complexity index(Casali et al. 2013), such measures might be adapted to distinguish the PP without binding, low-level perceptual binding, high-level binding involving actions of attention and imagination, and full procedural processing—the generation of a train of conscious thoughts modulated by initial high-level binding. As we have discussed, a broader definition of consciousness content includes the nonunified low-level binding occurring within each sensory modality (aligned with the recurrent processing ToC), while a less inclusive definition includes only the causes bound into the unified episode structure (aligned with the global workspace ToC). Action triggering and thought generation via imagination are additional processes that rely on binding but are possibly reduced or absent in some conscious states, such as meditation.